\documentstyle[preprint,tighten,aps]{revtex}
\def\lapprox{\mathrel{\mathop
  {\hbox{\lower0.5ex\hbox{$\sim$}\kern-0.8em\lower-0.7ex\hbox{$<$}}}}}
\def\gapprox{\mathrel{\mathop
  {\hbox{\lower0.5ex\hbox{$\sim$}\kern-0.8em\lower-0.7ex\hbox{$>$}}}}}
\def\mathrm{\mbox}
\def\eg{{\em e.g.}\/}
\def\etal{{\em et al.}\/}

\def\ie{{\em i.e.}\/}
\def\epsi{\epsilon}
\def\csi{\xi}

\def\fibe{\Phi_{\mbox{\footnotesize{Be}}}}

\newcommand{\unitan}{$\cdot 10^9 $cm$^{-2}$s$^{-1}$}

\newcommand{\sBe}{$^7$Be}

\begin{document}
\preprint{\null\hfill  INFNFE-08-96}
\title{The $^{51}$Cr neutrino source and Borexino: a desirable marriage}

\author{ N. Ferrari$^{1,2}$\footnote{nferrari@vaxmi.mi.infn.it},
         G.~Fiorentini$^{3,4}$\footnote{fiorentini@vaxfe.fe.infn.it} 
        and B.~Ricci$^{4}$\footnote{ricci@vaxfe.fe.infn.it}
}

\address{
$^{1}$Dipartimento di Fisica dell'Universit\`a di Milano, Via Celoria 16,
 I-20133-Milano,
Italy\\
$^{2}$Istituto Nazionale di Fisica Nucleare,  Sezione di Milano, Via 
Celoria 16,
       I-20133-Milano, Italy\\
$^{3}$Dipartimento di Fisica dell'Universit\`a di Ferrara,
       via Paradiso 12, I-44100 Ferrara, Italy\\
$^{4}$Istituto Nazionale di Fisica Nucleare, Sezione di Ferrara,
      via Paradiso 12, I-44100 Ferrara, Italy 
      }

%\date{}

\maketitle                 % Produces the title.

\begin{abstract}
Exposure to a $^{51}$Cr neutrino source as that used 
in Gallex will provide an excellent overall 
performance test of Borexino, which should collect 
about 1400 source induced events, with an 
initial rate of about 35 counts per day. This will 
be  particularly important if MSW-small-angle  
turns out to be the solution of  the solar neutrino
 problem. In addition, if an independent, 
accurate calibration is available, one will have  
an interesting experiment on neutrino properties:
 as an  example, a neutrino magnetic 
moment of the order $5\cdot10^{-11}\mu_B$
could be detected/excluded  at the  90\% C.L.
\end{abstract}

\vskip1.5truecm

Borexino at Gran Sasso \cite{Borexino} 
has a lot  to tell about  \sBe~solar 
neutrinos: Standard Solar Models (SSMs) predict a rate 
$\lambda_{\odot}\simeq 50$ counts per day  (c.p.d.), mostly 
from the \sBe~neutrinos, over an estimated background 
$\lambda_b \simeq 10$ c.p.d. At such 
rates, seasonal modulations of  $\lambda_{\odot}$, 
corresponding  to variations in the earth-sun distance  
$R_{ES}$, should allow a clean discrimination of signal  to background.  
In addition, the Just-So oscillation 
mechanism  predicts large seasonal modulations of the signal, well in 
excess of the $1/R_{ES}^2$
law, which are  clearly detectable with Borexino.

On the other hand, several scenarios predict much 
smaller event rates, comparable to or  even smaller 
than the expected bacgkround: this is the 
case of the MSW-small-angle   solutions, both for 
oscillations into active and sterile neutrinos. 
Also, if one insists on the  massless  neutrinos of the minimal
 electroweak theory, the available 
results (from Gallex \cite{Gallex}, Sage \cite{Sage}  Chlorine 
\cite{Davis} and Kamiokande \cite{Kamioka}) together 
with the luminosity constraint imply 
an extreme reduction of the \sBe~neutrino flux 
$\fibe$ with respect to  $\fibe^{SSM}$, see \eg~\cite{Bere,taup95noi}. 
A reasonable question is thus the following one:
\begin{itemize}
\item
a) How to assess definitely the capability of Borexino to detect 
$\nu_e$ from \sBe, if the 
solar flux is low?
\end{itemize}

In this case, an  overall performance test, with the
  exposure of Borexino to a man made neutrino source, is 
a must. We should like to demonstrate
  that  the  $^{51}$Cr source which was used by  the Gallex 
collaboration in their pioneer experiment \cite{calibration} 
looks ideal in this respect.

In addition, we remark that 
if  an independent,  accurate calibration of  Borexino is available,  the  neutrino 
source experiment   is interesting for the study of neutrino properties:
  the $\nu_e-e$ cross section can be 
accurately measured,  \eg~searching  for the contribution of a neutrino 
magnetic moment.
Also, neutrino 
oscillations along  the path from the source to the interaction point can be investigated 
(disappearance experiment). We consider thus the following questions:
\begin{itemize}
\item
b) What is the sensititvity to a neutrino magnetic moment?
\item
c) Which region in the oscillation plane can be investigated?
\end{itemize}

As a working hypothesis, we assume that the source is placed just ouside the external wall of 
the detector, at a distance $D=9$ m from the center (see fig.\ref{schema}), 
\ie~the easiest location. At the end 
of this note we shall consider the advantages/disadvantages of other 
possible configurations.\\

{\em
a)The overall performance test}

The relevance of a $^{51}$Cr  source as that which was used in Gallex,
with activity at end of bombardment  (EOB)
$A_{EOB}= 61.9\pm1.2$ PBq, is immediately understood by oberving that 
the corresponding  neutrino flux at the centre 
of the detector, $\phi_{EOB} = (6.1\pm0.1)$\unitan  is  quite close to 
the  solar
\sBe~neutrino flux predicted by 
SSMs: $\fibe^{SSM}$=5.1\unitan, according to Ref. \cite{BP95}.   As well 
known \cite{calibration},  the spectrum of the $^{51}$Cr source 
(see table \ref{tabCr}) is quite 
similar to that of the \sBe~electron capture,  so that the  source 
induced event rate $\lambda_s$ over its 
lifetime ($\tau=39.97\pm0.01$ days) is well comparable to the event rate 
predicted by SSMs. More precisely, one has:
\begin{equation}
\label{ratesource}
\lambda_s =  N_e  \langle \sigma _{CC+NC} \rangle \phi_s \, f \quad ,
\end{equation}
where full detection efficiency is assumed,  $N_e= 3.2\cdot 10^{31}$ is the number of electrons in the fiducial 
(100 tons)  scintillator mass and $\langle \sigma _{CC+NC} \rangle$ is the 
electroweak cross section of  $\nu_e-e$ 
scattering for kinetic energy of the recoiling electron 
T$_e\geq 250$ keV \cite{Borexino}, averaged 
over the source components (see table \ref{tabCr});
 $\phi_s$  is the neutrino flux at the detector center  and  
$f(R/D)$ is the geometrical factor accounting for the finite distance $D$ from the (pointlike) 
source to the center of a spherical detector with radius $R$ (for the present case 
$R=3$ m):
\begin{equation}
\label{fattore}
f(x)= \frac{3}{2x^3} \left [ x- \frac{1-x^2}{2} \, 
{\mbox{ln}} \left ( \frac{1+x}{1-x} \right )  \right ]    \quad .
\end{equation} 
All this gives, at EOB, $\lambda_s^{EOB} = 39$ c.p.d. Actually, it takes  
a few days to bring the source from the reactor to LNGS.  Assuming that, 
as in Gallex,  the exposure can start  about 4 days after EOB and 
that it extends over  a few lifetimes, one should collect a number of 
source induced events $N_s \simeq  90\% 
\lambda_s^{EOB}\tau =1400$.

There are two main differences with respect to Gallex:
i) a gain by an order of magnitude in statistics, originating mainly from the 
larger number of scattering centers. 
ii) As Borexino is a real time detector, the time dependence of source induced events can be 
accurately followed, so that background will be easily subtracted.

In the absence of background  and/or solar neutrino events, $N_s$ 
could be ultimately determined to the level of its statistical accuracy.
Here and in the following all uncertainties will be evaluated at the 90\% 
Confidence Level (C.L.). In this way one has 
 $\Delta N_s/N_s=1.64/\sqrt{N_s} =4.4\% $.
Actually, from the background and/or 
the sun  one expects a counting rate in the range (10--60) c.p.d. 
Correspondingly, we estimate an accuracy $\Delta N_s/N_s$ in the range
(6--10)\% from
Monte Carlo simulations.
We shall take
 $\Delta N_s/N_s= 8\%$ 
 as an indicative value, corresponding to $(\lambda_\odot 
 +\lambda_b)=30$ c.p.d.
  By summing in 
quadratures this uncertainty with that on the source activity
($\Delta A/A=3.1\%$),  the relative error 
in the comparison between expected and detected events will be, approximately:
\begin{equation}
\label{epsi}
\epsi= \sqrt{(\Delta N_s/N_s)^2 + (\Delta A/A)^2} = 8.6 \%  \quad  
 {\mbox{at 90\% C.L.  }} 
\end{equation}

All this means that, as one cannot switch the Sun  on and off,
 then it is best  to switch  another sun on.\\

{\em b)The source experiment and the neutrino magnetic moment }

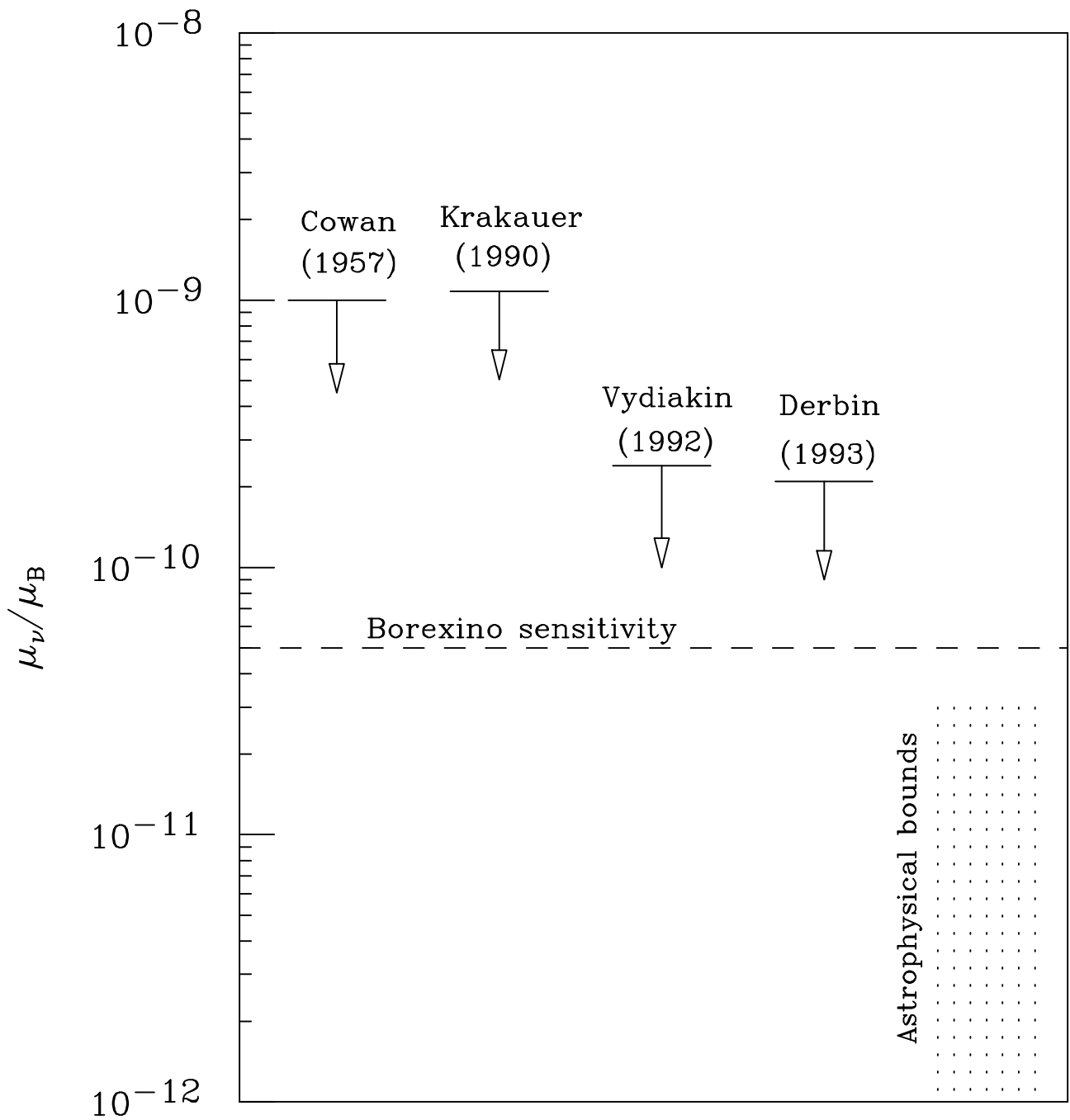
\begin{figure}
\caption[d]{
Upper bounds on $\mu_\nu$ from reactor \cite{Reines,Vydm,Derbin}
and accelerator experiments \cite{Krak}, at the 90\% C.L. (arrows),
 sensitivity of 
Borexino at the same C.L. (dashed line). Astrophysical 
and cosmological arguments suggest $\mu_\nu \leq 3\cdot 10^{-11} \mu_B$, 
see \eg~Ref. \cite{Raffelt}.
}
\label{magmom}
\end{figure}

If the detection efficiency is accurately determined,  the source 
experiment provides a precise measurement of the $\nu_e-e$ elastic cross 
section and the possible contribution of a neutrino magnetic moment 
$(\sigma _{CC+NC} \rightarrow \sigma _{CC+NC} + \sigma _{mm})$ is detectable.
The relative contribution to the cross section for 
$T_e>250$ keV, $\csi= \langle \sigma _{mm} \rangle / \langle 
\sigma _{CC+NC} \rangle $, is proportional to $\mu_\nu^2$ \cite{Kyul}
and one has $\csi = 0.39$ for  
$\mu_{\nu}=10^{-10} \mu_B$. 
One can thus study a neutrino magnetic moment: 
\begin{equation}
\label{mu}
			\mu_{\nu}/(10^{-10}\mu_B) = 
\sqrt{\epsi/\csi} \simeq 0.5 \quad  
 {\mbox{at 90\% C.L.  ,}} 
\end{equation}
about an order of magnitude better than  
given  from available experiments, close to the astrophysical 
upper-bounds (see fig. \ref{magmom}) and in the same range as planned in 
future experiments \cite{Rita,Carlo}.\\

{\em c)The source experiment and  neutrino oscillations }

Similar to the case of reactor experiments, one can perform a disappearance experiment 
\cite{Bahcalloscilla}, the main differences being: 
i) one is working  with neutrinos (not antineutrinos); 
ii) the energy  spectrum is precisely determined;  
 iii) detection is through 
$\nu-e$ scattering.

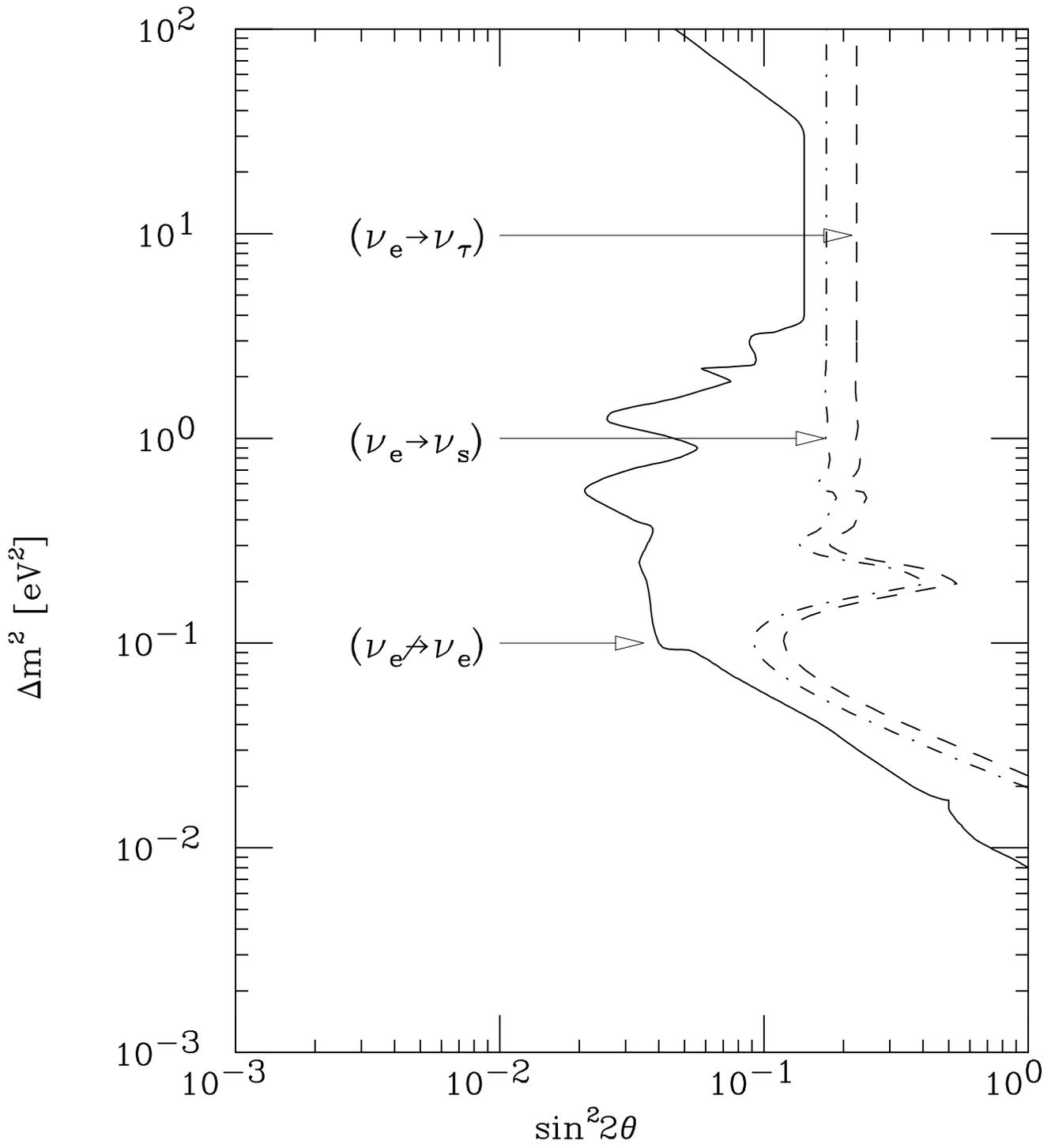
\begin{figure}
\caption[c]{
The  borders  of the 90\% C.L. regions explorable  with   
the source  at $D=9$ m from the 
detector center for oscillations into sterile neutrinos 
(dot-dashed line)  and into $\nu_\tau$ (dashed line).
 Available  results from reactor \cite{Bugey,Kurciatov} and accelerator 
 \cite{Bebc}
  experiments are also 
 shown (solid line).
}
\label{isocurve}
\end{figure}

Neutrino oscillation would result in a reduced number of detected events. For the case of 
oscillation into sterile neutrinos,  the  limiting (\ie~90\% C.L.) detectable transition 
probability $P_{\nu_e\rightarrow \nu_s}$, 
averaged over neutrino paths and energies,   is immediately obtained from 
the condition
$P_{\nu_e\rightarrow \nu_s}= \epsi$.
The explorable region in the ($sin^22\theta$, $\Delta m^2$) plane,  is shown  
in fig. \ref{isocurve}.

The reader can immediately verify the  intersections of the
 curve with the  borders of the figure, 
since for large mass differences $sin^22\theta =2 \epsi$ whereas for maximal mixing 
$\delta m^2 =4 \sqrt{\epsi} E / L$, where $L$ is the average of 
the squared path-lenghts and $E$ is a suitable average neutrino energy, 
essentially the energy of the most intense line.

Due to the neutral current (NC) interaction, 
the experiment is somehow less sensitive to oscillations into active 
neutrinos, say $\nu_{\tau}$,  see again fig. \ref{isocurve}.
With respect to the previous case,  one has   $sin^22\theta \rightarrow  
sin^22\theta (1- \rho)$ where  
$\rho= \langle \sigma _{NC} \rangle / \langle \sigma _{CC+NC} \rangle = 0.2 $, 
 which results in  a  uniform shift to the right in the logarithmic 
plot.

From fig. \ref{isocurve} one sees that the region which can be accessed 
with the source experiment is already excluded.
 This result is not so bad, as it actually provides an additional 
 test for Borexino: if the number of source events comes out smaller than 
 expected, it cannot be due to neutrino oscillations.\\

{\em d)The optimal  location of the source}

The source might be placed just outside the inner steel sphere, ad a 
distance $d=6.5$ m from the detector center.
With respect to the previous case, the event number $N_s$ increases by a 
factor 2; for $(\lambda_\odot + \lambda_b)=30$ c.p.d. the uncertainty is 
$\Delta N_s/ N_s=4.5\%$, comparable to $\Delta A /A$. In place of eq. 
(\ref{epsi}) one has now $\epsilon '=5.5\%$ A small gain
($\sqrt{\epsilon '/\epsilon }$) is obtained for the limiting  magnetic 
moment. Concerning neutrino oscillations, for $\Delta m^2 
\geq 1$ eV$^2$ one can reach $sin^2 2\theta =0.1$, a region
not completely excluded. On the other hand, as a consequence of the 
shorter baseline, the lower limit for $\Delta m^2$ increases for a factor
$D/d  \sqrt{\epsilon '/\epsilon}  \simeq 1.1$.

If the source is placed at the
 center of the detector (the hardiest choice from the technical point of 
 view), the expected rate for standard neutrinos  
[$\lambda_c= 
A \langle \sigma _{CC+NC} \rangle 3 N_e  /(4\pi R^2)$] is a factor 27
 larger than  when the source is placed just outside the tank.
However, with such a huge statistics, the error on the activity 
is now  dominant. 

In conclusion the easiest location (just outside the tank) looks as the 
most convenient one.\\

In the Borexino  proposal, the  exposure   to an intense beta emitter 
such as $^{90}$Sr   was considered and the potential for the study of 
antineutrino properties, as the magnetic moment, was discussed 
\cite{Borexino}.
The $^{51}$Cr source has several advantages: 
i) {\underline {neutrinos}}  with almost the same spectrum as the solar ones will be 
detected; ii) the detection reaction is the same for solar and source 
neutrinos and iii) last not not 
least,  the source is there (it has to be reactivated, of course). 
 We should  like to encourage our colleagues to exploit its potential, when Borexino 
will be ready. 

~\\
\centerline{****}

This analysis started from an interesting remark by E. Fiorini at the Gran Sasso meeting on
New Trends in Solar Neutrino Physics, May 2--4 1996.
We appreciated the stimulating atmosphere created there by V. Berezinsky.

We are grateful to G. Bellini, M. G. Giammarchi, S. Magni and M. Moretti for useful 
discussions.

%B. Ricci thanks ``Consiglio Nazionale delle Ricerche'' (CNR) for a  fellowship.

\begin{table}
\caption[aa]{
For each of  the four lines of the  $^{51}$Cr spectrum we show the 
neutrino energy  $E_\nu$, the relative intensity P, 
the  $\nu_e-e$ cross section $\sigma _{CC+NC}$, that for 
  $\nu_\tau-e$ scattering  $\sigma _{NC}$ (both calculated following
   ref. \cite{Bahcallsezioni})
and the contribution to the cross sections for a magnetic moment 
$\mu=10^{-10}\mu_B$.  All these have been 
integrated  for electron kinetic energies higher than 250 keV.  
In the last row, the values  averaged over the 
source spectrum are shown.
}

\begin{tabular}{lr@{}lr@{}lr@{}lr@{}lr@{}lr@{}lr@{}l}
%\hline
%\hline
         &\multicolumn{2}{c}{$ E_\nu $}
             &\multicolumn{2}{c}{P} 
                    &\multicolumn{2}{l}{$\sigma_{CC+CN}$}
                      &\multicolumn{2}{l}{$\sigma_{CN}$}
			&\multicolumn{2}{l}{$ \sigma_{mm}$} \\			
         &\multicolumn{2}{c}{[MeV]}
		 &\multicolumn{2}{c}{}
                    &\multicolumn{2}{l}{[10$^{-46}$  cm$^2$]}
			 &\multicolumn{2}{l}{[10$^{-46}$  cm$^2$]}
			 &\multicolumn{2}{l}{[10$^{-46}$  cm$^2$]} \\
\hline
         & 0. & 751 &    0.& 09 &    25.& 2  &  5.& 84   &  9.& 82 \\
         & 0. & 746 &    0.& 81 &    24.& 8  &  5.& 76   &  9.& 70 \\
         & 0. & 431 &    0.& 01 &    1. & 45  &  0.& 45   &  0.& 78 \\
         & 0. & 426 &    0.& 09 &    1. & 14  &  0.& 36   &  0.& 62 \\
\hline
         &    &     &  Aver&age:&    22.& 5  &  5.& 23   &  8. & 81
%\hline
%\hline
\end{tabular}
\label{tabCr}
\end{table}

\newpage
 
 \begin{figure}
 \caption[b]{
Schematic layout of Borexino and  the $^{51}$Cr source.}
 \label{schema}
\end{figure}

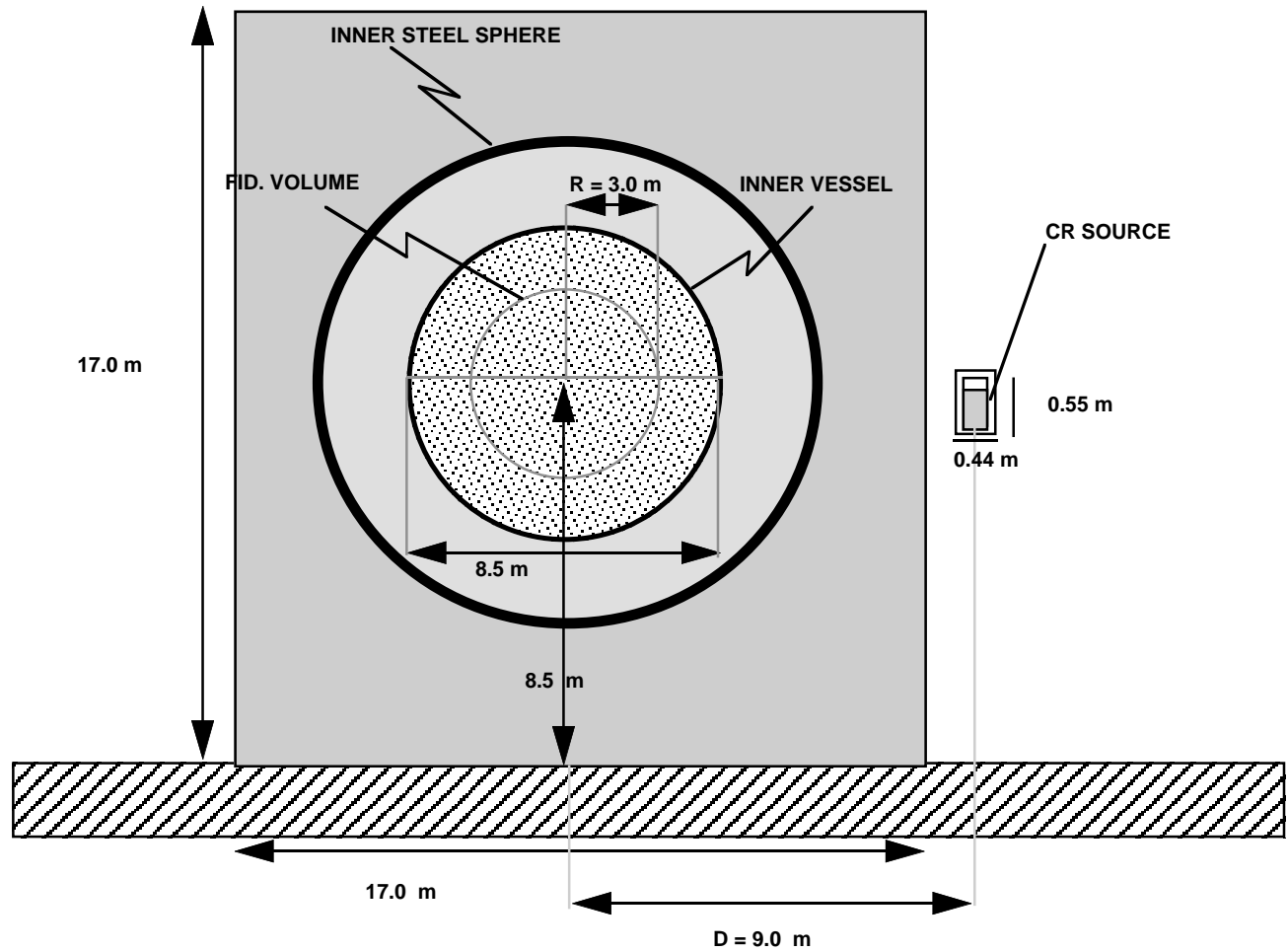
\end{document}